\documentclass[preprint,12pt]{elsarticle}
\usepackage{geometry}[0.8]
\usepackage{graphicx} 
\graphicspath{ {./images/} }
\usepackage[utf8]{inputenc}
\usepackage{wrapfig}
\usepackage[english]{babel}
\usepackage{enumerate}
\usepackage{amssymb}
\usepackage{amsthm}
\usepackage{mathtools}
\usepackage{bm}
\usepackage{import}
\usepackage{todonotes}
\usepackage{hyperref}
\usepackage{caption}

\newcommand*{\surf}{\mathcal{S}}
\newcommand*{\para}{\boldsymbol{X}}

\newcommand*{\height}{\xi}
\newcommand*{\velo}{\mathbf{V}}
\newcommand*{\veloT}{\mathbf{v}}
\newcommand*{\veloN}{v_{\perp}}
\newcommand*{\testVol}{\phi_{\text{Vol}}}
\newcommand*{\upd}{\mathbf{Y}}
\newcommand*{\testUpd}{\mathbf{Z}}
\newcommand*{\testMean}{\mathcal{G}}
\newcommand*{\gradC}{\nabla_{\!C}}
\newcommand*{\gradS}{\nabla_{\!S}}
\newcommand*{\divC}{\text{div}_{\!C}}
\newcommand*{\divS}{\text{div}_{\!S}}
\newcommand*{\DivC}{\text{Div}_{\!C}}
\newcommand*{\Dtm}{\text{D}_t^\mathfrak{m}}

\newcommand*{\gradFull}{\nabla}

\newcommand*{\lapC}{\Delta_C}
\newcommand*{\lapS}{\Delta_S}
\newcommand*{\press}{\lambda_{p}}
\newcommand*{\auxiliary}{\mu}
\newcommand*{\testAuxiliary}{\omega}
\newcommand*{\energy}[1][]{\mathcal{F}_{#1}}
\newcommand*{\D}[1][]{\bm{D}_{#1}}
\newcommand*{\diss}[1][]{\mathcal{R}\IfValueT{#1}{_{\text{#1}}}}
\newcommand*{\con}[1][]{\diss[#1]}

\newcommand*{\deformTensor}{\boldsymbol{\sigma}}
\newcommand*{\meanCurv}{\mathcal{H}}
\newcommand*{\gaussCurv}{\mathcal{K}}
\newcommand{\shapeOp}{\mathbf{I\!I}}
\newcommand*{\bendStiff}{\kappa}
\newcommand*{\visc}{\eta}
\newcommand*{\bend}{\boldsymbol{f}}

\newcommand*{\testVelo}{\bm{W}}
\newcommand*{\testHeight}{\mathfrak{h}}
\newcommand*{\testPress}{q}

\newcommand*{\idR}[1][]{{\boldsymbol{I}_{#1}}}
\newcommand*{\idS}[1][]{{\boldsymbol{P}_{#1}}}
\newcommand*{\normal}{\boldsymbol{\nu}}
\newcommand*{\TRS}[1][]{\mathbf{T}^{#1}\mathbb{R}^3\vert_{\surf}}

\newcommand*{\TS}[1][]{\mathbf{T}^{#1}\surf}

\newcommand*{\heightparamI}{a}
\newcommand*{\mobiH}{M_\height}
\newcommand*{\heightparamII}{b}
\newcommand*{\Truess}[1]{{\mathring{#1}}}
\newcommand*{\deform}{\eth_{\testVelo}}
\newcommand*{\LGrad}[2]{\left\langle#1,#2\right\rangle}
\newcommand*{\metric}{g}
\newcommand*{\sint}[1]{\int_\surf#1\,d\surf}
\newcommand*{\friction}{\gamma}
\newcommand*{\lagrVol}{\lambda_{\text{Vol}}}
\newcommand*{\lagrHeight}{p}
\newcommand*{\lagrNormal}{\lambda_{\nu}}
\newcommand*{\cell}{T}
\newcommand*{\discr}{h}
\newcommand*{\innerApprox}[2]{\inner{\discr}{#1}{#2}}
\newcommand*{\inner}[3]{\left(#2,#3\right)_{#1}}
\newcommand*{\err}[1]{e_{#1}}
\newcommand*{\spacetimenorm}[1]{\|{#1}_\discr - #1\|_{L^\infty(L^2(\surf_{\discr}))}}

\newcommand*{\veloDer}{\mathfrak{B}_{\testVelo}}

\begin{document}

\begin{frontmatter}



\title{Fluid deformable surfaces with variable thickness - a Surface Shallow-Water-Helfrich model} 


\author[label1]{Jan Magnus Sischka}
\affiliation[label1]{organization={Institut f\"ur wissenschaftliches Rechnen, TU Dresden},
            city={Dresden},
            postcode={01062}, 
            country={Germany}}

\author[label1,label2,label3]{Axel Voigt}
\affiliation[label2]{organization={Center for Systems Biology Dresden (CSBD)},
            addressline={Pfotenhauerstra{\ss}e 108}, 
            city={Dresden},
            postcode={01307}, 
            country={Germany}}
\affiliation[label3]{organization={Cluster of Excellence Physics of Life (PoL)},
            city={Dresden},
            postcode={01062}, 
            country={Germany}}

\begin{abstract}
Epithelial tissues play a fundamental role in morphogenesis. Mechanically they can be viewed as thin soft materials exhibiting a solid-fluid duality. The Surface Navier-Stokes-Helfrich model accounts for these properties by combining bending
and surface hydrodynamics. In order to account for varying cell thickness we incorporate a thickness field in the spirit of a shallow-water equation, but defined on the (self-)evolving
surface. This replaces the inextensibility constraint of the two-dimensional fluid and
with it the conservation of surface area by an incompressibility constraint of the thin film fluid allowing for changes in surface area. We develop a numerical scheme based on surface finite elements, perform convergence tests, demonstrate the impact on shape evolution of closed surfaces with a constant enclosed volume and discuss implications on modeling morphogenesis.
\end{abstract}




\end{frontmatter}

  \section{Introduction}

Fluid deformable surfaces are ubiquitous interfaces in biology, playing an
essential role in processes from the subcellular to the tissue scale. From a mechanical point of view, fluid deformable surfaces are soft materials exhibiting a solid–fluid duality: while they store elastic energy when stretched or bent, like solid shells, they cannot do so under in-plane shear, a situation
under which they flow as two-dimensional, viscous fluids. This leads to a tight coupling between tangent flows and shape changes in the presence of curvature. Any shape change contributes to strains and so must
be accompanied by a tangent flow and vice versa, tangential flows on curved surfaces
induce shape deformations. This coupling makes curvature a natural element that shapes cellular organization and tissue function. The basic model describing this behavior is a Surface Navier-Stokes-Helfrich model \cite{arroyo2009relaxation,torres2019modelling,reuther2020numerical,nitschke2026hydrodynamic}. It combines bending and surface hydrodynamics and assumes the two-dimensional fluid to be inextensible. While mainly applied to model the shape evolution of lipid membranes \cite{torres2019modelling,reuther2020numerical,sahu2025arbitrary,nitschke2026hydrodynamic}, there are also attempts to use this model on larger scales to describe the evolution of epithelial tissues. Two notable examples are shape instabilities resulting from additional chiral contributions to the stress \cite{AA23,PV24}, which follow from surface active nematic forces \cite{NV24}, and local growth with three-dimensional confinement, which leads to invagination \cite{PBV25} and can be related to the basic phenomena of gastrulation. However, in these applications the assumption of a surface with zero thickness is probably oversimplified. The cells in the epithelial tissue have a finite thickness. While in most cases this thickness is significantly smaller than any lateral scale, there is strong evidence that the thickness is not constant. Experiments on epithelial monolayers on curved substrates indicate that cells exhibit a pronounced thickening on regions with negative curvature, whereas cells appear less thick on regions with positive curvature \cite{luciano2021cell,luciano2026multiscale}. Assuming the cell volume to be fixed, this change in thickness must be accompanied by a reduction or widening of cell area. Indeed such an anti-correlation between cell area and tissue thickness has been suggested in \cite{luciano2026multiscale}. Similar effects are also known for in vivo processes. One example is the intensively studied process of gastrulation in {\em Drosophila}. In \cite{rauzi2015embryo} tissue mechanics of this process is addressed and cell shapes are analyzed, demonstrating that, e.g., dorsal cell widening, which is a largely passive process, is compensated by shortening their thickness. 

We aim to incorporate a thickness field in the Surface Navier-Stokes-Helfrich model in the spirit of a shallow-water equation, but defined on a (self-)evolving surface. The thickness/height is thereby measured with respect to the normal of the surface. This replaces the inextensibility constraint of the two-dimensional fluid and with it the conservation of surface area by an incompressibility constraint of the thin film fluid and with it conservation of tissue volume. This modification maintains the solid-fluid duality and with it the tight coupling between tangent flow and shape changes in the presence of curvature, but enhance the evolution by an additional degree of freedom, which allows for area changes as a result of thickness variations. The thickness field enters as an additional scalar surface quantity. We also consider the bending rigidity to depend on the thickness/height following the proposed scaling in \cite{friesecke2006hierarchy} for solid thin structures. The equations are derived using the Lagrange-d'Alembert principle \cite{Marsden_2010} following the derivations of two-phase fluid deformable surfaces \cite{bachini2023derivation} and Surface Beris-Edwards-Helfrich models \cite{nitschke2025beris,nitschke2026liquidcrystallineorderimpact}. The resulting model is, with respect to geometric coupling terms, nonlinearities and the structure of the equations, similar to the Surface Cahn-Hilliard-Navier-Stokes-Helfrich model considered in \cite{bachini2023derivation,sischka2025two}. As in previous studies we restrict the investigation to closed surfaces and as in \cite{krause2023numerical,nestler2023stability,KV24,PV24,sischka2025two,garcke2025parametric,PBV25} we also consider a constraint on the enclosed volume, which makes the problem nonlocal. 

The computational approach is based on isoparametric surface finite elements (SFEM) \cite{DE13,nestler2019finite}, an Arbitrary Lagrangian-Eulerian (ALE) approach \cite{reuther2020numerical,sauer2025curvilinear,sahu2025arbitrary}, a 'BGN' type method to maintain mesh regularity \cite{Barrett_SIAMJSC_2008}, classical Lagrange elements and a semi-implicit time discretization. For the special case of a constant thickness profile the equations reduce to the classical Surface Navier-Stokes-Helfrich model and the numerical scheme reduces to the one considered in \cite{krause2023numerical}, which has been analyzed in \cite{garcke2025parametric}. It extends proposed relations for Stokes flow on stationary surfaces between the polynomial degree of the Lagrange elements to the degree for the surface parameterization \cite{brandner2022finite,hardering2025parametric,reusken2025analysis} and experimentally leads to the same optimal order also for the Surface Navier-Stokes-Helfrich model.

The paper is structured as follows: In Section \ref{sec:Model} we introduce the notation, formulate the full set of equations, and derive them using the Lagrange-d'Alembert principle, which guarantees thermodynamic consistency. Furthermore we discuss two special cases, the classical Surface Navier-Stokes-Helfrich model, which follows by constraining the thickness to be constant, and the Surface Shallow-Water model, which follows by constraining the normal velocity to vanish. In Section \ref{sec:Numerics} we 
describe the considered numerical approaches for the full and the constraint models. In Section \ref{sec:Results} we explore the main features of the model, with a strong focus on parameter regimes corresponding to the envisioned applications in tissue evolution. Finally we draw conclusions in Section \ref{sec:Conclusions}. In \ref{app:Derivations} we provide details on the variational derivatives considered in the model derivations.

  \section{Model} \label{sec:Model}
  \subsection{Notation}
  We mainly follow the notation considered in \cite{bachini2023derivation,nitschke2025beris}. We consider a time dependent smooth and oriented surface $\surf = \surf(t)$ without boundary, embedded in $\mathbb{R}^3$ with parametrization $\para=\para(t, y_1, y_2)$. We denote by $\normal$ the outward pointing surface normal, the surface projection is $\idS=\idR[3]-\normal\otimes\normal$, with $\idR[3]$ the identity matrix in $\mathbb{R}^3$, the shape operator is $\shapeOp = -\gradC\normal$, the mean curvature $\meanCurv= \operatorname{tr}\shapeOp$, where $\operatorname{tr}$ is the trace operator, and the Gaussian curvature $\gaussCurv = 1/2 (\meanCurv^2 - \|\shapeOp\|^2)$. We consider time-dependent Euclidean-based $ n $-tensor fields denoted as $\TRS[n]$. We call $\TRS[0] = \TS[0]$ the space of scalar fields on $\surf$, $\TRS[1]=\TRS$ the space of vector fields on $\surf$, and $\TRS[2] $ the space of 2-tensor fields on $\surf$. Important subtensor fields are tangential tensor fields $\TS[n]\leq\TRS[n]$.
    

  Let $\height \in \TS[0]$ be a continuously differentiable scalar field (thickness), $\velo \in \TRS$ a continuously differentiable $\mathbb{R}^3$-vector field (velocity), and $\deformTensor \in \TRS[2]$ a continuously differentiable $\mathbb{R}^{3\times3}$-tensor field (strain rate tensor), defined on $\surf$. We define the componentwise surface gradient by $\gradC\height = \gradFull\height^e$, $\gradC\velo = \gradFull\velo^e\idS$ and $\gradC\deformTensor = \gradFull\deformTensor^e \idS$, where $\height^e$, $\velo^e$ and $\deformTensor^e$ are arbitrary smooth extensions of $\height$, $\velo$ and $\deformTensor$ in the normal direction and $\gradFull$ is the gradient of the embedding space $\mathbb{R}^3$. The trace of those operators give rise to one notion of divergence operators, which for vector fields $\velo$ and 2-tensor fields $\deformTensor$ are defined as $\DivC\velo = \operatorname{tr}(\gradC\velo)$ and $\DivC(\deformTensor) = \operatorname{tr}_{(2,3)}\gradC(\deformTensor)$. The relations to the covariant derivatives $\gradS$ and the covariant divergence $\divS$ on $\surf$, with $\lapS = \divS \cdot \gradS$ the Laplace-Beltrami operator, read $\gradC\height=\gradS\height$ and
  ${\DivC\velo = \divS(\idS\velo)-(\velo\cdot\normal)\meanCurv}$, respectively. The surface velocity $\velo = \veloT + \veloN \normal$ can be split into its tangential part $\veloT=\idS \velo$ and its normal part $\veloN = \velo\cdot\normal$. Considering this splitting we can also write $\gradC \velo = \gradS \veloT - \veloN \shapeOp + \normal \otimes (\gradS \veloN + \shapeOp \veloT)$ and ${\DivC\velo = \divS \veloT - \veloN \meanCurv}$. 
  
  We consider $\Truess{\height} = \Dtm\height + \height \DivC\velo$ the scalar Truesdell time derivative of $\height$ \cite{backofen2026scalar}, with $\Dtm \height = \partial_t \height+ (\veloT - \veloT_0)\cdot\nabla_\surf \height$ the material time derivative for $\veloT_0$ the observer velocity. For details on different time derivatives on evolving surfaces we refer to \cite{CERMELLI_FRIED_GURTIN_2005,DE13,nitschke2023tangential}.
  
  \subsection{Governing equations}\label{sec:governing}
  The governing equations which couple the shape evolution, the surface velocity and the thickness in strong form read: \\
  
  \noindent Find the surface $\surf$, the surface velocity $\velo$, the surface thickness $\height$, and the Lagrange multiplier $\lagrVol$ such that
  \begin{align}
    \height\Dtm \velo =& \; \visc \DivC\!\left(\height\deformTensor\right) - \friction\velo -\lagrVol\normal + \bend\label{eq:moments}\\
    \Truess{\height} =& \;0\label{eq:conti}\\
    \int_\surf \velo \cdot \normal\,d\surf=& \;0 \label{eq:volume}
  \end{align}
  with force
  \begin{align}\label{eq:forces}
      \bend = &-\DivC\left(\heightparamI\gradS\height\otimes\gradS\height-\left(\heightparamI\height\lapS\height-\heightparamII(\height-\height_0)\height+\frac{\heightparamI}{2} \|\gradS \height\|^ 2 + \frac{\heightparamII}{2}(\height-\height_0)^2\right)\idS\right)\nonumber\\
      &-\left(\bendStiff(\height)\meanCurv\|\shapeOp\|^2 - \frac{\bendStiff(\height)-\bendStiff'(\height)\height}{2}\meanCurv^3
         + \lapS(\bendStiff(\height)\meanCurv)\right)\normal \nonumber\\
      &- \frac{\bendStiff''(\height)\height\gradS\height}{2}\meanCurv^2 - \bendStiff'(\height)\height\meanCurv\gradS\meanCurv,
  \end{align}  
  and given initial conditions for $\surf(0)$, $\velo(0)$ and $\height(0)$.
  Here, $\deformTensor = \frac{1}{2}\left(\idS\gradC\velo + (\idS\gradC\velo)^T\right)$ is the strain rate tensor, $\visc > 0$ is the surface viscosity, $\friction \geq 0$ a friction parameter, $\heightparamI > 0$ and $\heightparamII > 0$ denote parameters, $\height_0 > 0$ is a desired thickness and $\bendStiff(\height) > 0$ is the bending rigidity. For more details we refer to the derivation of these equations in Section \ref{sec:Model Derivation} and \ref{app:Derivations}. The equations are defined on the evolving closed surface $\surf(t)$ and are supplemented by initial conditions for $\surf(0)$, $\velo(0)$ and $\height(0)$.
  
  \subsection{Model derivation using the Lagrange d'Alembert principle} \label{sec:Model Derivation}

  The derivation of the governing equations is based on the integral Lagrange-d'Alembert principle. A theoretical foundation for this approach is provided by \cite{Marsden_2010} and it has been proven to be a powerful tool to derive thermodynamically consistent models for two-phase surface deformable surfaces \cite{bachini2023derivation} and Surface Beris-Ewards-Helfrich models \cite{nitschke2025beris,nitschke2026liquidcrystallineorderimpact}. In both cases it provides the basis for a detailed investigation of the strong interplay of surface curvature, surface hydrodynamics and surface phase composition or surface liquid crystal ordering, respectively. We here follow the same general setting, which requires an action functional ${\cal{A}} = \int_{\cal{T}} {\cal{L}} dt $, the time integral of the Lagrangian $ {\cal{L}} = \energy[K] - \energy[P]$, comprising the kinetic energy $\energy[K]$ and a (free) potential energy $\energy[P]$, and an energy flux potential ${\cal{R}}$. The integral Lagrange-d'Alembert principle results from equating variations of the action functional ${\cal{A}}$ with variations of the time integral of the flux potential $\int_{\cal{T}} {\cal{R}} dt$. The variation of the action functional ${\cal{A}}$ is given by the perturbation of the state variables by virtual displacements and the variation of the time integral of the flux potential $ \int_{\cal{T}} {\cal{R}} dt $ is given by perturbing the process variables by virtual displacements. The primary state variables include the parameterization $\para$ in $\mathbb{R}^3$ and the thickness field $\height \in \TRS[0]$. The main process variables are the material velocity field $\velo\in\TRS $ and a thickness field rate $\Truess{\height} \in \TRS[0]$. Furthermore, one can consider state and process variables in the form of Lagrange functions or Lagrange multipliers to enforce constraints on both the primary state and process variables. In our case we only enforce constraints on the process variables, which are summarized in the set ${\cal{C}}_{\cal{R}}$, which contains $\lambda_\gamma \in \TRS[0]$ and $\lagrVol \in \mathbb{R}$. The integral Lagrange-d'Alembert principle reads
  \begin{align}
  &\left( \frac{\delta {\cal{A}}}{\delta \para}, \mathbf{W} \right)_{L^2({\cal{T}};\TRS)} + \left( \frac{\delta {\cal{A}}}{\delta \height}, \psi \right)_{L^2({\cal{T}};\TRS[0])} \nonumber \\ &= \int_{\cal{T}} \left( \frac{\delta {\cal{R}}}{\delta \velo}, \mathbf{W}\right)_{L^2(\TRS)} + \left( \frac{\delta {\cal{R}}}{\delta \Truess{\height}}, \psi \right)_{L^2(\TRS[0])} + \sum_{\gamma \in {{\cal{C}}_{\cal{R}}}} \left( \frac{\delta {\cal{R}}}{\delta \lambda_\gamma}, \phi_\gamma \right)_{L^2(\TRS[0])}  \nonumber\\ 
  & \qquad + \inner{L^2(\TRS[0])}{\frac{\delta \diss}{\delta\lagrVol}}{\testVol} dt 
  \label{eq:LDA_time_global}
  \end{align}
  for all virtual displacements $\mathbf{W} \in \TRS$, $\psi \in \TRS[0]$, $\phi_\gamma \in \TRS[0]$, $\testVol\in \mathbb{R}$. The localization in time is achieved through the arbitrariness of the time interval ${\cal{T}}$, and the localization in space through the arbitrariness of the virtual displacements and lead after defining $\energy[P]$, $\energy[K]$, ${\cal{R}}$ and ${\cal{C}}_{\cal{R}}$ to the governing equations \eqref{eq:moments} - \eqref{eq:volume}.

  We consider the (free) potential energy $\energy[P] = \energy[T] + \energy[H]$ with 
  \begin{align}
    \energy[T] =& \int_{\surf}\frac{\heightparamI}{2}\|\gradS\height\|^2 + \frac{\heightparamII}{2}(\height-\height_0)^ 2\,d\surf\,, \label{eq:FT}\\
    \energy[H] =& \int_{\surf} \frac{\bendStiff(\height)}{2}\meanCurv^2 \, d\surf\,, \label{eq:FH}
  \end{align}
  with $a>0$ and $b>0$ parameters, $\height_0>0$ a reference thickness and $\kappa(\height) > 0$ a thickness dependent bending rigidity. $\energy[T]$ penalizes variations in $\height$ and deviations from a reference thickness. This provides a minimal energy to enforce a smooth thickness profile. $\energy[H]$ is a classical Willmore/Helfrich energy, neglecting spontaneous curvature and Gaussian bending rigidity. 

  The kinetic energy is defined as 
    \begin{align}
    \energy[K] =& \int_{\surf}\frac{\height}{2}\|\velo\|^2\,d\surf\,.\label{eq:FK}
  \end{align}
  
  The energy flux potential is considered as $\diss = \diss[V] + \diss[R] + {\diss}_{\height} + \diss[p] + \diss[Vol]$. We address viscous dissipation, friction with the surrounding medium and immobility with respect to the thickness, and assume an incompressible fluid in the thin film and conservation of the enclosed volume. The corresponding dissipation potentials read
  \begin{align}
    \diss[V] =& \int_{\surf} \visc\height\|\deformTensor\|^2\,d\surf\,,\\
    \diss[R] =& \int_{\surf} \frac{\friction}{2}\|\velo\|^2\,d\surf\,,\\
    {\diss}_{\height} =& \int_{\surf} \frac{\mobiH}{2}\Truess{\height}^2\,d\surf\,,
  \end{align}
  where $\visc$ is the surface viscosity, $\friction$ a friction parameter, and $\mobiH$ an immobility parameter. 

\begin{wrapfigure}{r}{0.43\textwidth}
    \centering
    \includegraphics[width=0.43\columnwidth]{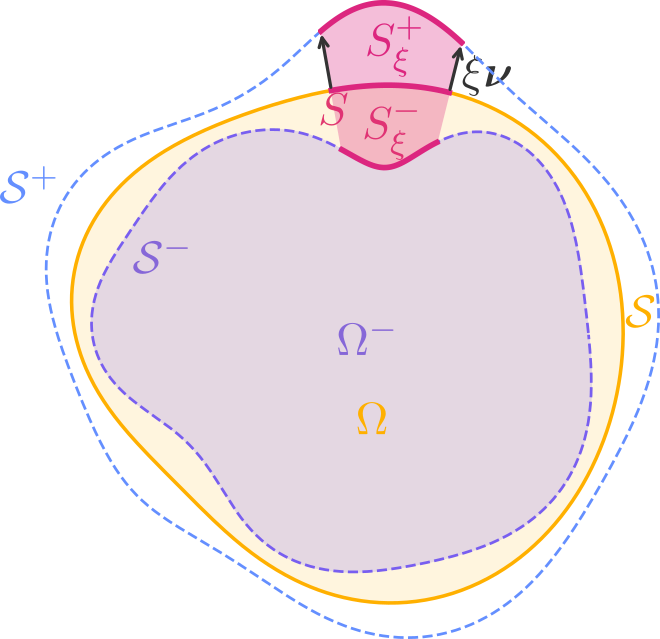}
    \caption{Sketch of the geometric quantities. Shown is a cut through the surface $\surf$ in yellow and the negatively and positively displaced surfaces $\surf^-$ and $\surf^+$ in purple and blue, respectively. Cuts through the inner enclosed volume $\Omega^-$ and the volume $\Omega$ enclosed by $\surf$ are shown by purple and yellow shading, respectively. A subset $S$ of $\surf$ and the volumes $S_\height^-$ and $S_\height^+$ are shown in pink. Exemplary outer normals, scaled by the thickness $\height$ are shown in black.} 
    \label{fig:sketch_surface}
  \end{wrapfigure}

  In addition we assume an incompressible fluid in the thin film and conservation of the enclosed volume. Let $S\subset \surf$ an arbitrary subset of $\surf$ and $S_\height^\pm =\{\para(t, y_1, y_2) \pm r\height(\para(t, y_1, y_2))\normal : \para(t, y_1, y_2)\in S, r\in[0,1]\}$ the normal extension of $S$ in either positive or negative direction, and $S_\height = S_\height^- \cup S_\height^+$. See Fig. \ref{fig:sketch_surface} for a sketch.

  Up to second order, the volume of $S_\height^\pm$ reads \cite{zhong1989bending}
  \begin{align}
      \int_{S_\height^\pm}\,d\Omega &= \int_S \height \mp \height^2\meanCurv\,d\surf
  \end{align}

  Note that for constant thickness $\height$ the volume can be computed exactly by Weyl's tube formula \cite{weyl1939volume, gray2003tubes}
  \begin{equation}\label{eq:Weyl}
     \int_{S_\height^\pm}\,d\Omega = \int_S \height \mp \height^2 \meanCurv + \frac{\height^3}{3} \gaussCurv\, d\surf\,.
  \end{equation}
  By the assumption that variations in normal direction are negligible, incompressibility of the thin film fluid reduces to 
  \begin{equation}
      0 = \Dtm\int_{S_\height}\,d\Omega = \Dtm\int_S 2\height\,d\surf +\mathcal{O}(\height^3) \,.
  \end{equation}
  Assuming that $\height$ is small, we neglect higher order contributions, leading to the well known relationship \cite{arroyo2009relaxation, torres2019modelling,backofen2026scalar}
  \begin{equation} \label{eq:con_incomp}
      \Dtm\int_{S}2\height\,d\surf = \int_S2\Truess{\height}\,d\surf = 0\,.
  \end{equation}
  We therefore define the constraint 
  \begin{align}
      \con[p] &= \int_\surf \press\Truess{\height}\,d\surf\,,\label{eq:con_p}
  \end{align}
  with Lagrange function $\press\in\TS[0]$. For the conservation of enclosed volume, we consider $\Omega^-$ the volume enclosed by the inner surface $\surf^- = \{\para(t, y_1, y_2) -\height(\para(t, y_1, y_2))\normal\}$. It holds
  \begin{align}
      \Dtm\int_{\Omega^-} \,d\Omega = \Dtm\int_\Omega \,d\Omega - \Dtm\int_{\surf_\xi^-}\,d\Omega &= \int_\surf \frac{1}{3}\velo \cdot \normal\,d\surf - \Dtm\int_\surf \height + \height^2 \meanCurv\, d\surf +\mathcal{O}(\height^3) \nonumber \\
      &\approx \int_\surf \frac{1}{3}\velo \cdot \normal\,d\surf - \int_\surf \Truess{\height}\, d\surf \,,
  \end{align}
  where $\Omega$ is the volume enclosed by $\surf$ and in the last step we neglect higher order contributions with respect to $\height$. To ensure $\Dtm \int_\Omega^- \,d\Omega = 0$ and using eq. \eqref{eq:con_incomp} we thus specify the constraint
  \begin{align}
      \con[Vol] &= \lagrVol \int_\surf \velo \cdot \normal\,d\surf\, ,\label{eq:con_v}
  \end{align}
  with Lagrange multiplier $\lagrVol\in\mathbb{R}$. 
  
  This specifies $\energy[P]$, $\energy[K]$, ${\cal{R}}$ and ${\cal{C}}_{\cal{R}}$ and eq. \eqref{eq:LDA_time_global} leads after localization in time the governing equations in weak form
  \begin{align}\label{eq:Lagrange-DAlembert}
      0&=\left(\height(\Dtm\velo) + \D[\para]\energy + \D[\velo]\diss,\testVelo\right)+ \left(\D[\height]\energy +\D[\Truess{\height}]\diss ,\testHeight \right)\nonumber\\
      &\hspace{12em}+ \left(\D[\press]\con[p] , \testPress\right) + \left(\D[\lagrVol]\con[V], \testVol\right).
  \end{align}
  Where $\D[\bullet]$ is the variational derivative with respect to $\bullet$. The derivation and the specific forms of all terms are provided in \ref{app:Derivations}. Using them and localizing in space leads to the strong formulation eq. \eqref{eq:moments} - \eqref{eq:volume}. 
  
  As a result of the construction we obtain
  \begin{equation}
      \frac{d}{dt}\left(\energy[K] + \energy[P]\right) = - 2 {\cal{R}} \leq 0\; ,
  \end{equation}
  which demonstrates thermodynamic consistency. 
  
  \subsection{Model simplifications}\label{sec:simplified}

  In the following we link the proposed governing equations to simplified models already discussed in the literature. We consider two special cases.

  \subsubsection{Surface Navier-Stokes-Helfrich model}\label{sec:Surface_NSH}

  We consider a constant thickness profile, which we enforce using an additional constraint 
  \begin{equation}
      {\diss}_\height = \sint{\lagrHeight \Dtm\height}\,,
  \end{equation}
  where $\lagrHeight \in\TS[0]$ is a Lagrange function. Variation with respect to $\velo$ leads to the additional term $\gradS\left(\lagrHeight\height\right) + \lagrHeight\height\meanCurv\normal$
  which is added to the right hand side of eq \eqref{eq:conti}. Variation with respect to $\lagrHeight$ leads to the additional constraint 
  \begin{align}\label{eq:con_height}
      \Dtm\height = 0\,,
  \end{align} 
   which together with eq. \eqref{eq:conti} leads to the inextensibility constraint $ \DivC\velo = 0$. All other variations do not contribute additional terms to the system. For a detailed derivation see \ref{app:Derivations}. Assuming the initial thickness $\height(0) = 1$, eqs. \eqref{eq:moments} - \eqref{eq:volume} thus simplify to: \\

   \noindent Find the surface $\surf$, the surface velocity $\velo$, the surface pressure $p$ and the Lagrange multiplier $\lagrVol$ such that
  \begin{align}
    \Dtm \velo =& \; -\gradS\lagrHeight - \lagrHeight \meanCurv\normal + \visc \DivC \deformTensor - \friction \velo  - \lagrVol\normal + \bend\label{eq:moments_1}\\
    \DivC\velo =& \;0\label{eq:conti_1}\\
    \int_\surf \velo \cdot \normal\,d\surf=& \;0 \label{eq:volume_1}
  \end{align}
  with force
  \begin{align}
      \bend = &- \bendStiff \left(\meanCurv \left( \|\shapeOp\|^2 - \frac{1}{2}\meanCurv^2 \right)
         + \lapS \meanCurv \right)\normal\,,
  \end{align}
  defined on $\surf(t)$, given initial conditions for $\surf(0)$ and $\velo(0)$.
  
  This is the classical Surface Navier-Stokes-Helfrich model for an inextensible surface fluid. This model has been considered numerically in~\cite{reuther2020numerical,krause2023numerical,sauer2025curvilinear,sauer2025curvilinear,garcke2025parametric} and in the Stokes-limit in \cite{torres2019modelling,PBV25,igel2026streamfunctionformulationsurfacestokeshelfrich}. For analytical results see \cite{WaZZ12}. For further numerical and analytical approaches under additional symmetry assumptions we refer to \cite{al2020shear}, where a linear stability analysis of a tube-geometry is considered, and to \cite{olshanskii2023equilibrium}, where potential rotational symmetric equilibrium configurations are addressed. For a comparison of different derivations of the model we refer to \cite{reuther2015interplay,reuther2018erratum} and \cite{brandner2022derivations}. In the overdamped limit $\gamma \to \infty$ the classical Willmore/Helfrich flow is obtained with inextensibility and volume constraint, see \cite{bachini2023derivation}, and a huge literature exists on numerical approaches for the resulting system, see, e.g. \cite{du2004phase,Barrett_SIAMJSC_2008,elliott2010modeling,aland2014diffuse}.
  
  \subsubsection{Shallow-water equation on stationary surface}
  \label{sec:shallow-water}
  We assume a stationary surface $\surf$ by enforcing an additional constraint on the normal velocity 
  \begin{equation}\label{eq:con_normal}
      {\diss}_{\normal} = \sint{\lagrNormal\velo\cdot\normal}\,,
  \end{equation}
  where $\lagrNormal\in\TS[0]$ is a Lagrange function acting as a process variable. 
  
  The variation of \eqref{eq:con_normal} with respect to the velocity $\velo$ leads to the term $\lagrNormal\normal$, which enters the right hand side of eq. \eqref{eq:moments} as a purely normal term. Variation with respect to $\lagrNormal$ leads to the additional equation $\velo \cdot \normal = 0$. 

  Projecting eqs.~\eqref{eq:moments} - \eqref{eq:volume} onto the surface restricts the system of equations to a purely tangential problem. Considering eq. \eqref{eq:moments}, we follow the approach in \cite{jankuhn2018incompressible,reuther2018erratum,reuther2020numerical}. We use the identity 
  \begin{equation}
      \idS \Dtm \velo = \idS \Dtm \veloT + \idS\Dtm (\veloN \normal) = \Dtm \veloT + \Dtm \veloN \normal + \veloN \Dtm \normal = \Dtm \veloT\,,
  \end{equation} 
  to project the left hand side of eq. \eqref{eq:moments}. The tangential part of the modified right hand side reads
  \begin{align}
      \eta\idS\DivC(\height\deformTensor) - \gamma\idS\velo + \idS\bend - \lagrNormal\idS \normal =  \visc \divS\!\left(\height\tilde\deformTensor\right) -\friction\veloT+ \tilde\bend,
  \end{align}
  with rewritten strain rate tensor $\tilde{\deformTensor} = \frac{1}{2}\left(\gradS\veloT + (\gradS\veloT)^T\right)$ and modified force   \begin{align}
      \tilde\bend =&-\heightparamI\divS\left(\gradS\height\otimes\gradS\height\right)-\gradS\left(\heightparamI\height\lapS\height-\heightparamII(\height-\height_0)\height+\frac{\heightparamI}{2} \|\gradS \height\|^ 2 + \frac{\heightparamII}{2}(\height-\height_0)^2\right)\nonumber\\
      &- \frac{\bendStiff''(\height)\height\gradS\height}{2}\meanCurv^2 - \bendStiff'(\height)\height\meanCurv\gradS\meanCurv .
  \end{align}
  Note that the Lagrange function $\lagrNormal$ does not appear in the projected system of equations. Thus, we can eliminate the corresponding equation $\velo \cdot \normal = 0$ from the system and obtain
  \begin{align}
    \height\Dtm \veloT =& \; \visc \divS\!\left(\height\tilde\deformTensor\right) -\friction\veloT+ \tilde\bend\label{eq:moments_sw}\\
    \Dtm \height + \height\divS\veloT =& \;0\label{eq:conti_sw}
  \end{align}
  where eq. \eqref{eq:conti_sw} is the projected version of eq. \eqref{eq:conti}. These equations are defined on $\surf$ and supplemented by initial conditions $\veloT(0)$ and $\xi(0)$.
  
  With a different definition for $\tilde\bend$, eqs. \eqref{eq:moments_sw} and \eqref{eq:conti_sw} resemble the shallow-water equations with viscous terms in non-conservative form on a curved surface. The equations are most commonly considered in flat space or on a sphere
  to model geophysical flows, see, e.g. \cite{tan1992shallow} for a review. While for these applications the hyperbolic version without dissipation is most suited, also the viscous version of the  equations have been considered, see, e.g. \cite{kloeden1985global, wang2005cauchy}. The general derivation of the shallow-water equations is in accordance with our assumption on the film thickness to be much smaller than the lateral length scale. This together with the incompressibility of the thin film fluid implies that the tangential velocity field $\veloT$ is constant throughout the depth of the fluid. Integrating the film in normal direction allows the normal velocity within the film to be removed from the equations. This basic derivation of the shallow-water equations also holds on a curved surface and can even be extended to evolving surfaces. Examples with changing topographies of the surface can be found in \cite{tort2014,abreu2023geometrically}.

  While the connection to these equations is appealing, the numerical challenges for the intended use in geophysical flow, e.g., to handle shocks and contact discontinuities, do not apply for our application. On the length scale of the considered epithelial tissue the viscous term dominates and the equations can be handled with standard surface finite element methods (SFEM) \cite{DE13,nestler2019finite}.
  
  \section{Numerical method} \label{sec:Numerics}
  We consider a surface finite element method (SFEM) \cite{DE13,nestler2019finite} to solve the highly nonlinear set of geometric and surface partial differential equations \eqref{eq:moments}-\eqref{eq:forces} by combining the system with a mesh redistribution approach, know as the 'BGN' method, see \cite{Barrett_SIAMJSC_2008}. \cite{nestler2019finite} can be considered as an extension of \cite{DE13} from scalar-valued to vector- and tensor-valued surface partial differential equations, following an approach independently proposed in \cite{jankuhn2018incompressible,nestler2018orientational,hansbo2020analysis}. Following the approaches in \cite{bachini2023derivation,krause2023numerical,PBV25}, the additional equations for the parametrization are
  \begin{align}
    \partial_t \para \cdot \normal &= \velo\cdot\normal \label{eq:normalvel}\\
    \meanCurv \normal &= \lapC \para \,. \label{eq:meandiff}
  \end{align}
  These equations not only address the movement in normal direction but also generate a tangential mesh movement to maintain the shape regularity of the discretization and additionally provide an implicit representation of the mean curvature $\meanCurv$. In this sense, the method provides an Arbitrary Lagrangian-Eulerian (ALE) approach \cite{krause2023numerical,sauer2025curvilinear,sahu2025arbitrary,PBV25,garcke2025parametric}, Lagrangian in normal direction and Eulerian in tangential direction.
  
  We consider a discrete $k$-th order approximation $\surf_\discr^k$ of $\surf$, with $h$ denoting the discretization in space. We use the DUNECurvedGrid library \cite{praetorius2020dunecurvedgrid} and consider each geometrical quantity like the normal vector $\normal_\discr$, the shape operator $\shapeOp_\discr$, the Gaussian curvature $\gaussCurv_\discr$, and the inner products $(\cdot , \cdot)_\discr$ with respect to $\surf_\discr^k$. In the following we will drop the index $k$. We define the discrete function spaces for scalar functions by $V_{k}(\surf_\discr)=\{ \psi \in C^0(\surf_\discr) \vert\, \psi\vert_{\cell}\in\mathcal{P}_{k}(\cell)\}$ and for vector fields by $\boldsymbol{V}_{k}(\surf_\discr)=[V_{k}(\surf_\discr)]^3$. Within these definitions $\cell$ is the mesh element and $\mathcal{P}_{k}$ are the polynomials of order $k$. We consider $\velo_\discr,\para\in\boldsymbol{V}_2(\surf_\discr)$,  $\meanCurv_\discr,\height_\discr\in V_2(\surf_\discr)$, and introduce a new variable $\auxiliary_\discr\in V_2(\surf_\discr)$ to deal with the third derivative in the term $-\gradS\left(\heightparamI\lapS\height\right)$ appearing in equation \eqref{eq:forces} and a corresponding test function $\testAuxiliary_\discr \in V_2(\surf_\discr)$. This leads to an isogeometric setting. We discretize in time using constant time stepping with step size $\tau$. In each time step $t^n$ we solve the following system of equations: 
  \begin{align}  
    \frac{1}{\tau}\innerApprox{\height^{n-1}_\discr(\velo_\discr^n - \velo_\discr^{n-1})}{\testVelo_\discr} \hspace{-2em}&\hspace{2em}+ \innerApprox{\height^{n-1}_\discr(\veloT_\discr^{n-1} - \veloT_{0\discr}^{n-1})\cdot\gradC\velo_\discr^{n}}{\testVelo_\discr} \nonumber\\
    =& -\visc\innerApprox{\height^{n-1}_\discr\deformTensor\left(\velo_\discr^n\right)}{\gradC\testVelo_\discr} - \gamma\innerApprox{\velo^{n}_\discr}{\testVelo_\discr} \nonumber \\
    & + \heightparamI \innerApprox{\gradS\height^{n-1}_\discr \otimes \gradS\height^{n-1}_\discr - \height^{n}\auxiliary^{n-1}_\discr - \frac{1}{2}\|\gradS\height^{n-1}_\discr\|^2}{\DivC\testVelo_\discr} \nonumber \\
    & + \frac{\heightparamII}{2}\innerApprox{\height^{n-1}_\discr\height^n_\discr - \height_0^2}{\DivC\testVelo_\discr}  - \lambda^{n}_\discr \innerApprox{\normal^{n-1}_\discr}{\testVelo_\discr} \nonumber\\
    & - \innerApprox{\bendStiff(\height^{n-1}_\discr)\meanCurv^n_\discr\|\shapeOp^{n-1}_\discr\|^2 \normal_\discr^{n-1}}{\testVelo_\discr}\nonumber\\
    & + \innerApprox{\frac{\bendStiff(\height^{n-1}_\discr)-\bendStiff'(\height^{n-1}_\discr)\height^{n-1}}{2}\left(\meanCurv^{n-1}_\discr\right)^2\meanCurv^{n}_\discr\normal_\discr^{n-1}}
    {\testVelo_\discr} \nonumber \\
    & -\innerApprox{\lapS\left(\bendStiff(\height^{n-1}_\discr) \meanCurv^n_\discr\right) \normal^{n-1}_\discr} {\testVelo_\discr} \nonumber \\
    & -\innerApprox{\frac{\bendStiff''(\height^{n-1}_\discr)\height^{n-1}_\discr\gradS\height^{n}_\discr}{2}\left(\meanCurv^{n-1}_\discr\right)^2}{\testVelo_\discr}\nonumber\\
    & - \innerApprox{\bendStiff'(\height^{n}_\discr)\height^{n-1}_\discr\meanCurv^{n-1}_\discr \gradS\meanCurv^{n}_\discr}{\testVelo_\discr} \label{eq:moments_discr} \\
    \frac{1}{\tau}\innerApprox{\height^{n}_\discr - \height^{n-1}_\discr}{\testHeight_\discr} & + \innerApprox{(\veloT_\discr^{n-1} - \veloT_{0\discr}^{n-1})\cdot\gradC\height^{n}_\discr}{\testHeight_\discr} \nonumber \\
    =& - \innerApprox{\height^{n-1}_\discr\DivC\velo^n_\discr}{\testHeight_\discr} \label{eq:conti_discr} \\
          \innerApprox{\auxiliary^n_\discr}{\testAuxiliary_\discr} =& \innerApprox{-\gradS\height^n_\discr}{\gradS\testAuxiliary_\discr} \\
              \frac{1}{\tau}\innerApprox{\upd^{n}\cdot \normal^{n-1}_\discr}{\testMean_\discr} = & \innerApprox{\velo^n_\discr\cdot \normal^{n-1}_\discr}{\testMean_\discr} \\
    \innerApprox{\meanCurv^{n}_\discr \normal^{n-1}_\discr}{\testUpd_\discr} = & - \innerApprox{\gradC\left(\upd^n + \para^{n-1}_\discr\right) }{\gradC\testUpd_\discr}.
  \end{align}
  Note that the Lagrange multiplier $\lambda^{n}_\discr$ is unknown, which leads to an underdetermined system. To resolve that problem, we follow the approach introduced in \cite{krause2023numerical}. In order to fulfill the volume constraint, $\lambda^{n}_\discr$ has to be chosen such that 
  \begin{align}
    \Phi(\lambda^{n}_\discr)\coloneqq \int_{\surf^{n-1}_\discr} \velo^{n}_\discr(\lambda^{n}_\discr)\cdot \bm{\nu}^{n-1}_\discr\,d\surf = 0.
  \end{align}
  We consider $\Phi(\lambda) = 0$ as an equation in $\lambda$ and apply a Newton iteration $\lambda^{j+1} = \lambda^{j} - \Phi(\lambda^{j}) / \Phi'(\lambda^{j})$. Each iteration thus requires to solve the equations. After convergence is achieved the last step requires to compute the new surface $\surf^{n}_\discr$ by updating the paramerization $\para^{n} =  \para^{n-1} + \upd^{n}$, lifting the solutions $\velo^{n}_\discr$, $\height^{n}_\discr$, $\auxiliary^{n}_\discr$ and $\meanCurv^{n}_\discr$ to the new surface and to compute the remaining geometric quantities $\normal^{n}_\discr$, $\shapeOp^{n}_\discr$ and $\gaussCurv^{n}_\discr$ for the new surface.
  
  The discrete systems are implemented within the finite element toolbox AMDiS \citep{vey2007amdis,witkowski2015software} based on DUNE \citep{sander2020dune,alkamper2014dune}. 

  For the implementation of the model simplifications described in Section \ref{sec:simplified} we make the following adjustments to the system. For the Surface Navier--Stokes--Helfrich model, eqs. \eqref{eq:moments_1} - \eqref{eq:volume_1}, all terms depending on variations of $\height$ or the thickness dependent bending rigidity disappear and the right hand side of \eqref{eq:moments_discr} is extended by the term $-\innerApprox{{p}_\discr^{n}}{\DivC\testVelo_\discr}$. This leads to numerical issues related to the saddle-point structure of the system of equations, which is resolved by choosing $ {p}_\discr \in V_1(\surf_\discr)$ leading to $\mathcal{P}_{2}-\mathcal{P}_{1}$ Taylor--Hood elements for the surface velocity-pressure pair. The approach thus coincides with previous numerical methods \cite{KV24,PBV25,garcke2025parametric} and theoretical results \cite{hardering2025parametric}. Instead of implementing a new scheme for the shallow-water equations on a stationary surface, eqs. \eqref{eq:moments_sw} and \eqref{eq:conti_sw}, we follow the approach considered in \cite{jankuhn2018incompressible,reuther2020numerical,krause2023surface} and simply consider the full model but penalize the normal velocity by adding the term $(\alpha (\velo_\discr^{n} \cdot \normal_\discr^{n-1} - 0) \normal_\discr^{n-1}, \testVelo_\discr)$ to the right hand side of \eqref{eq:moments_discr}, with $\alpha > 0$ being a penalization parameter. A general error analysis for this approach is done in \cite{hardering2023tangential}.
  
  \section{Results} \label{sec:Results}
  \subsection{Considered setting and parameters}\label{sec:parameters}
  As initial surfaces $\surf(0)$, we either consider a sphere of radius $1$ centered at the origin or a sphere of radius $1$, perturbed by a spherical harmonic 
  \begin{align}
      \surf_0 &= \left\{r+r_0Y^m_l(\phi,\vartheta): \phi \in [0,\pi],\vartheta\in[-\pi,\pi] \right\}, \quad r_0>0, \nonumber\\
      Y^m_l(\phi,\vartheta) &= \sqrt{\frac{2l+1(l-m)!}{4\pi(l+m)!}}P^m_l(\cos\vartheta)e^{im\phi}.
  \end{align}
  with the Legendre polynomial $P^m_l$. We consider the cases $l = 3, m = 2, r = 1, r_0 = 0.5$ and $l=5, m=3, r=1, r_0=0.5$.
  For the initial thickness $\height(0)$ we either choose a constant thickness $\height(0) = 0.1$ or a two dimensional bump function $h_0 + c\exp\left(- \|(x_1, x_2)^T\| / (2\sigma)\right)$ with $h_0 = 0.1, c = 0.2, \sigma = 0.05$, projected on the sphere along the $z$ axis. The velocity field is always initialized with $\velo(0) = 0$.

  As material parameters we choose $\visc = 100$, $\friction = 1$. For the bending rigidity $\bendStiff(\height)$, we either choose constant bending rigidity $\kappa = 0.0667$ or thickness dependent bending rigidity $\kappa(\height) = \frac{\kappa}{3}\height^3$ with $\kappa = 200$, which for a constant thickness of $\height = 0.1$ leads to the same Helfrich energy.
  
  \subsection{Convergence study}
\begin{figure}[htbp]
  \centering
  \includegraphics[width=0.8\textwidth]{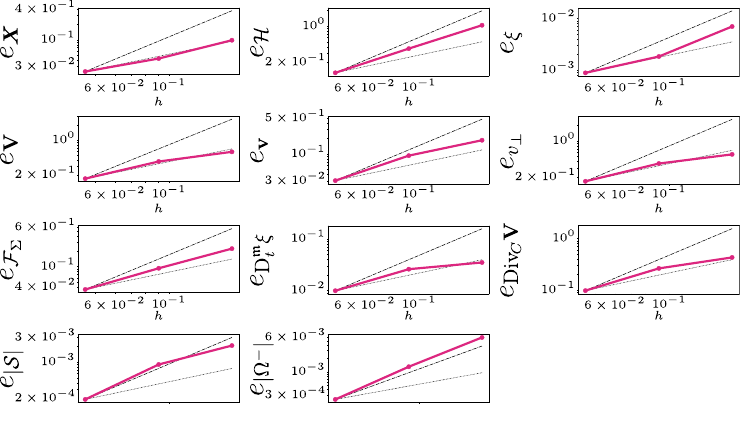}
    \caption{Convergence study for the full model. Shown are the space-time errors for the parameterization $\para$, the mean curvature $\meanCurv$, the thickness $\height$, the velocity $\velo$, the tangential velocity $\veloT$, the normal velocity $\veloN$, the energy ${\energy}_\Sigma$, the componentwise divergence $\DivC\velo$, the material time derivative of the thickness $\Dtm\height$, the surface area $|\surf|$ and the enclosed volume $|\Omega^-|$. Convergence orders of 1 and 2 are indicated by dotted and dashed lines, respectively.}
    \label{fig:panel_err}
  \end{figure}

  First, we perform convergence tests. As analytical results are not available, we consider experimental convergence studies against a numerical solution. As initial surface $\surf(0)$ we consider the perturbed sphere described in the previous section, with constant initial thickness $\height(0) = 0.1$, thickness dependent bending rigidity $\bendStiff(\height)$ and $\heightparamI = 0.1$, $\heightparamII = 100.0$. We run simulations until $t=1.0$ for four levels of refinements with $\discr = 0.180, 0.090, 0.045, 0.0225$. Thereby $\discr$ measures the largest edge length of the surface triangles on the initial surface. We also adapt the time step size $\tau$ accordingly to fulfill the CFL condition, leading to $\tau = 0.1, 1.25\times 10^{-2}, 1.5625\times 10^{-3}, 1.953125\times10^{-4}$. The solution with smallest $\discr$ is considered as reference solution. We consider a space-time error $L^\infty(L^2(\surf_h))$, denoting the $L^2$ error in space and the $L^\infty$ error in time, or only the $L^\infty$ error in time, i.e.
  \begin{align}
      \err{X} &= \spacetimenorm{\para}, &\err{\meanCurv} &= \spacetimenorm{\meanCurv}, \nonumber\\ \err{\height} &= \spacetimenorm{\height}, &\err{\velo} &= \spacetimenorm{\velo}, \nonumber\\ \err{\veloT} &= \spacetimenorm{\veloT}, & \err{\veloN} &= \spacetimenorm{\veloN}, \nonumber\\
      \err{{\energy}_{\Sigma}} &= \spacetimenorm{{\energy}_{\Sigma}}, & \err{\DivC\velo} &= \spacetimenorm{\DivC\velo}, \nonumber\\
      \err{\Dtm\height}&= \spacetimenorm{\Dtm\height}, & \err{|\surf|} &= \| |\surf_h| - |\surf| \|_{L^\infty(\surf_h)}, \nonumber\\
      \err{|\Omega^-|} &= \| |\Omega^-_h| - |\Omega^-| \|_{L^\infty(\surf_h)}
  \end{align}
  of the discrete solutions $\para_\discr$, $\velo_\discr$ and $\height_\discr$, and the resulting quantities $\meanCurv_\discr$, ${\veloT}_\discr$, ${\veloN}_\discr$, ${\energy}_{\Sigma \discr}$, $|\surf_\discr|$ and $|\Omega^-_\discr|$, where the last denote the area of $\surf_\discr$ and the enclosed volume of $\Omega_{\discr}^{-}$, for varying $\discr$, where $\para$, $\velo$, $\height$ and the corresponding resulting quantities $\meanCurv$, $\veloT$, $\veloN$, ${\energy}_{\Sigma}$, $|\surf|$ and $|\Omega^-|$, here denote the discrete reference solution for the finest $\discr$. The error plots are shown in Fig. \ref{fig:panel_err} and indicate at least first order convergence for all quantities, which is sufficient for our purpose to explore the physical implications of the model.

  \subsection{Physical implications of the model}
  With confidence in the numerical scheme we next perform numerical simulations to explore the physical implications and the coupling between tangent flow and shape changes in the presence of curvature, area changes and thickness variations. We consider this complex interplay step by step.  
  
  \subsubsection{Comparison to shallow water model on a stationary surface}
  We first consider the effect of thickness variations by comparing the full model to the shallow water model on a stationary surface. As initial surface we consider a unit sphere. The initial height profile is a gaussian bump as described in Section \ref{sec:parameters}. The considered parameters are $\heightparamI = 0$ and $\heightparamII = 2000$. In case of the full model we additionally consider $\kappa(\height) = \frac{200}{3}\height^3$, in case of the surface shallow water model we choose $\alpha = 500$. We run the simulations until time $t = 1.0$. For both models we consider two values for the surface viscosity, $\visc = 100$ and $\visc = 2$. The lower viscosity, is for demonstration only and leads to inertia effects like the formation of waves, while the larger value is more suited for the context of epithelial tissue. 

  \begin{figure}[htbp]
    \centering
    \includegraphics[width=0.9\textwidth]{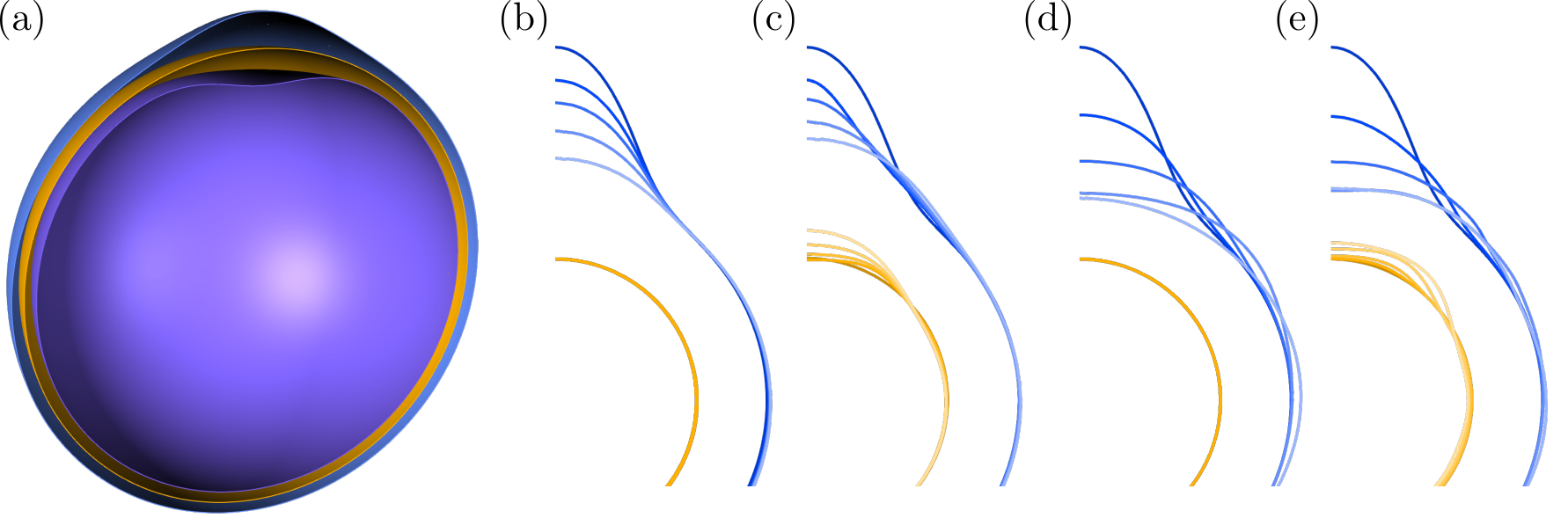}
  
    \caption{Comparison between the full model and the shallow-water model. (a) Cut through the initial shape. The initial surface $\surf(0)$ is shown in yellow, the surfaces displaced by the thickness profile $\surf_\height^+$ and $\surf_\height^-$ are shown in blue and in purple, respectively. (b) - (e) Equatorial cuts of the evolution of the surface $\surf$ in yellow and the displaced surface $\surf_\height^+$ in blue are shown. The displacement is magnified $10$ times for visualization. Shown are the evolutions for different values of the viscosity $\visc = 100$ ((b), (c)), and $\visc = 2$ ((d), (e)) and different time steps ranging from dark to light colors. In both cases, the shallow water model is shown on the left ((b), (d)) and the full model on the right ((c), (e)). For $\visc = 100$ the time steps are $t = 0, 0.05, 0.1, 0.2$ and $0.4$, for $\visc = 2$ the time steps $t = 0, 0.01, 0.02, 0.04$ and $0.08$ are shown.}
    \label{fig:overview_sw}
   \end{figure}

  \begin{figure}[htbp]
        \centering
        \includegraphics[width=0.9\textwidth]{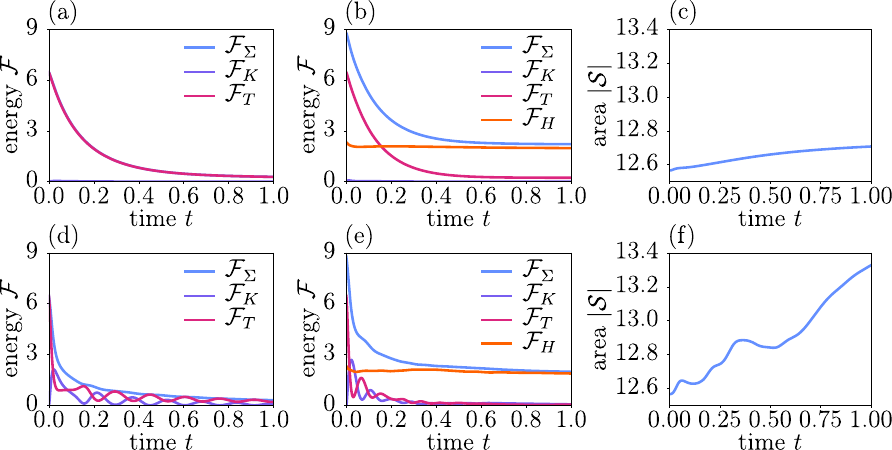}
    \caption{Evolution of the (free) potential and kinetic energies of the surface shallow-water model, (a) and (d), the full model, (b) and (e), and the evolution of the area $|\surf|$, (c) and (f), corresponding to the results in Fig. \ref{fig:overview_sw}. The top row, (a) - (c), shows the evolution for $\visc = 100$, the bottom row (d) - (f), for $\visc = 2$. The sum of the energies is given as $\energy[\Sigma] = \energy[K] +\energy[T] + \energy[H]$ for the full model, or $\energy[\Sigma] = \energy[K] + \energy[T]$ for the surface shallow-water model. In (a) the curves for $\energy[\Sigma]$ and $\energy[T]$ are indistinguishable}. 
    \label{fig:evo_energy}
\end{figure}

In Fig. \ref{fig:overview_sw} the evolution of the systems for the four parameter combinations between the full model, the shallow water model and the different choices for $\visc$ are shown at five time steps. Fig. \ref{fig:overview_sw} (a) shows the initial configuration, Fig. \ref{fig:overview_sw} (b) and (c) show the evolutions for $\visc = 100$ for the shallow water model and the full model, respectively, and Fig. \ref{fig:overview_sw} (d) and (e) show the evolutions for $\visc = 2$ for the shallow water model and the full model, respectively. For the last the expected properties of the shallow water equations, the relaxation of the bump and formation of waves are clearly visible. For the full model this behavior has an impact on the shape evolution, the surface moves outwards at the bump, propagating with the wave. For $\visc = 100$ this behavior is less pronounced. The bump is damped without the formation of waves. However, the impact on the shape remains.

This behavior is quantified in Fig. \ref{fig:evo_energy}, where we consider the evolution of the energy ${\energy}_\Sigma$ and its various components, as well as the evolution of $|\surf|$. The top row, (a) - (c), shows the quantities for the realistic case $\eta = 100$ and the bottom row, (d) - (f), for $\eta = 2$. All demonstrate the decrease in ${\energy}_\Sigma$ and the strong impact of $\energy[T]$ on ${\energy}_\Sigma$. For $\eta = 2$ the oscillation of $\energy[K]$ and $\energy[T]$ demonstrate the interplay between tangential flow and shape variations ($\energy[K]$) and variations in thickness ($\energy[T]$). In this setting $\energy[H]$ only plays a minor role, as $\surf$ remains close to a sphere. However, the intended change in surface area $|\surf|$ is clearly visible. For $\eta = 2$ the area increases, which is a result of the redistribution of the initial bump and the formation of a thin film layer with constant thickness. For $\eta = 100$ the impact is stronger and shows the coupling with the oscillations in $\energy[T]$. At $t =1.0$ the simulations did not yet converged to a stationary state. For the full model we expect this state to be the same for both values of $\eta$ and to correspond to a sphere with constant thickness $\xi$.

\subsubsection{Comparison to Surface Navier--Stokes--Helfrich model}
Next, we explore the influence of variations in the surface thickness $\height$ on the evolution of the system. Therefore, we compare the full model to the Surface Navier--Stokes--Helrich model. Due to incompressibility of the thin fluid layer, changes of the layer thickness lead to local growth and shrinkage of the surface area, resembling an extensible fluid surface. 

To investigate the qualitative changes of this phenomena, we perform simulations for both models, starting from a spherical surface, perturbed by a spherical harmonic with $l=5$ and $m=3$ as described in Section \ref{sec:parameters}. For the Surface Navier--Stokes--Helfrich model, we choose a constant bending rigidity $\kappa = 0.0667$, for the full model we consider $\kappa(\height) = \frac{200}{3}\height^3$. This leads to the same Helfrich energy $\energy[H]$ for $\height = 0.1$, which is the initial thickness $\height(0) = 0.1$ and reference height $\height_0$. Other parameters are $\heightparamI = 0$ and $\heightparamII = 2000$.

\begin{figure}[htbp]
    \centering
    \includegraphics[width=0.9\textwidth]{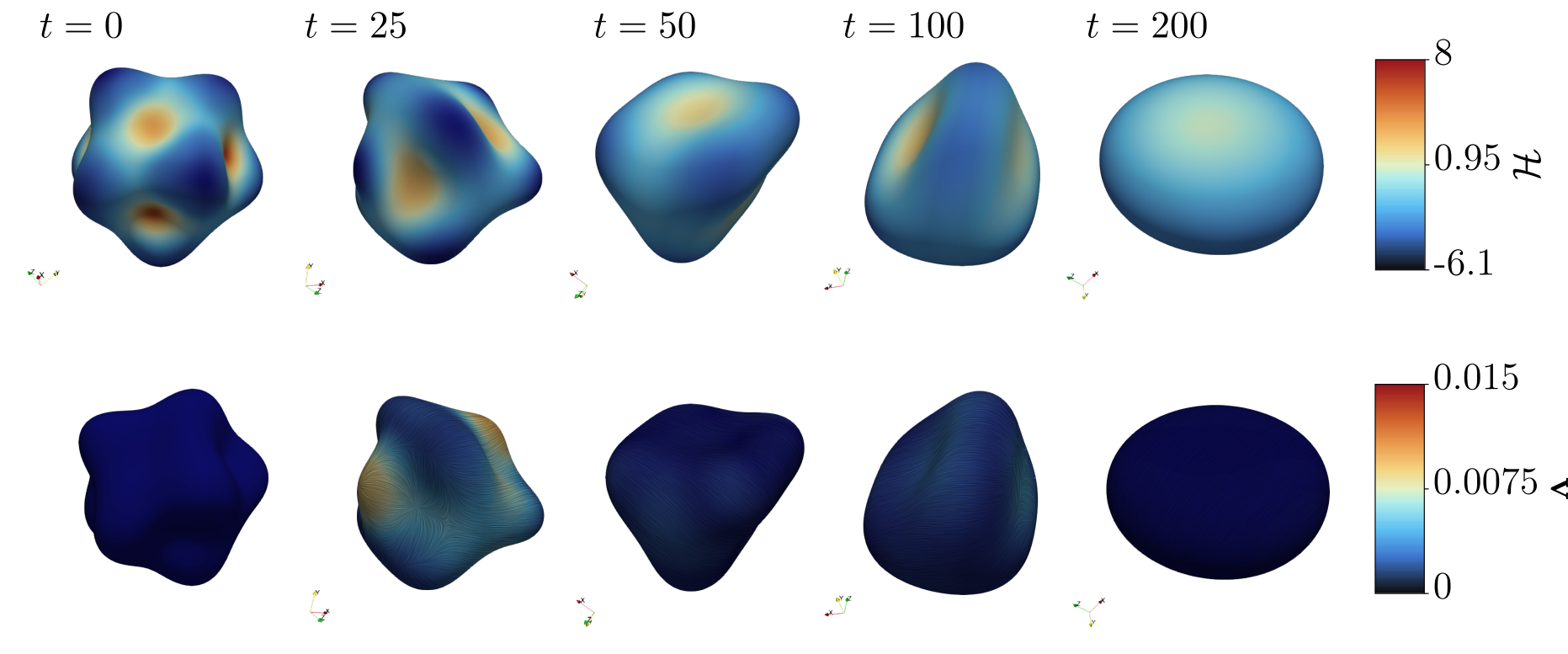}
    \captionof{figure}{Snapshots of the evolution for the Surface Navier--Stokes--Helfrich model as describesd in Section \ref{sec:Surface_NSH}. Shown in color are the evolution of the mean curvature $\meanCurv$ (top row) and the tangential velocity $\veloT$ (bottom row) using LIC (line intensity convolution). The timesteps shown are $t = 0, 25, 50, 100, 200$.}
    \label{fig:panel_no_thick}
    \vspace{1em}
    \centering
    \includegraphics[width=0.9\textwidth]{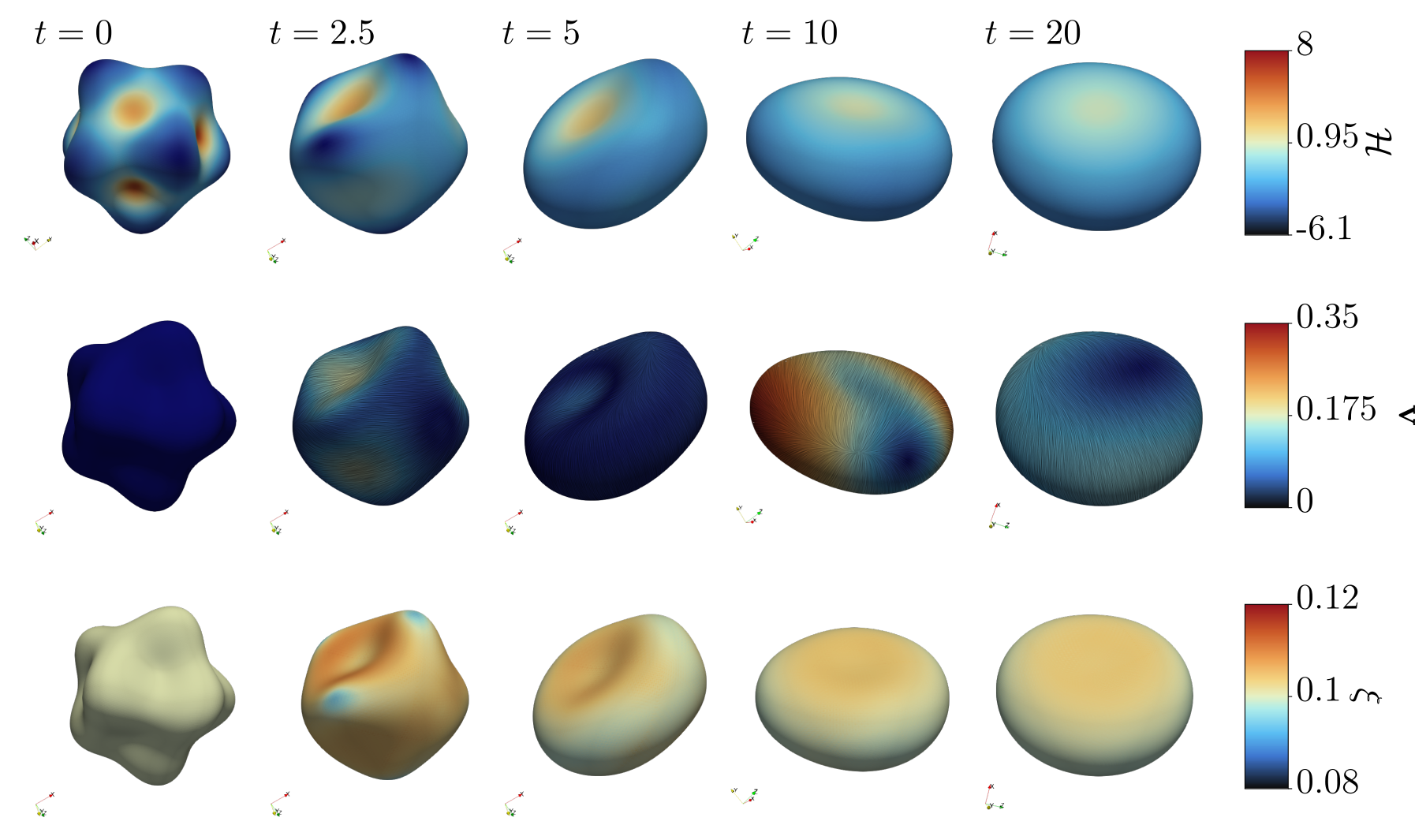}
    
    \captionof{figure}{Snapshots of the evolution for the full model. Shown in color are the evolution of the mean curvature $\meanCurv$ (top row), the tangential velocity $\veloT$ (middle row), using LIC, and the thickness $\height$ (bottom row). The timesteps shown are $t = 0, 2.5, 5, 10, 20$.}
    \label{fig:panel_thick}
\end{figure}

Fig. \ref{fig:panel_no_thick} shows the evolution for the Surface Navier-Stokes-Helfrich model and Fig. \ref{fig:panel_thick} the evolution of the full model. In both cases the shapes at selected time instances are shown together with the mean curvature $\meanCurv$ and the tangential velocity $\veloT$. For the full model in addition we show the surface thickness $\xi$. The simulations run until an equilibrium is reached. 

The time scale significantly differs, in the full model the equilibrium is reached approximately one order of magnitude faster. In both cases an oblate shape is reached. However, the evolutions significantly differ. The characteristic three-fold-like shape between $t=25$ and $t=100$ for the Surface Navier-Stokes-Helfrich model, is not present in the full model. Also the tangential velocity field differs, the magnitude is larger at some of the selected time instances for the full model. This change in morphology and flow field, and the enhanced evolution is a consequence of the additional degree of freedom to change the thickness. Thereby strongly negative curvature regions lead to reduced thickness. 

\begin{figure}
    \centering
    \includegraphics[width=0.9\textwidth]{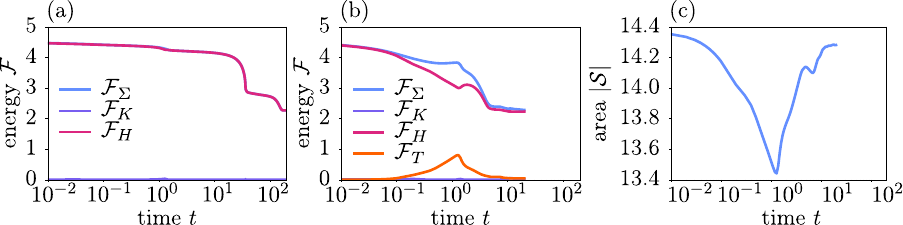}
    \caption{Evolution of the (free) potential and energies of (a) the Surface Navier--Stokes--Helfrich model, corresponding to Fig. \ref{fig:panel_no_thick} and (b) the full model, corresponding to Fig. \ref{fig:panel_thick}. (c) The evolution of the surface area $|\surf|$ corresponding to the results in  Fig. \ref{fig:panel_thick}. The sum of the energies is given as $\energy[\Sigma] = \energy[K] + \energy[H]$ for the Surface Navier--Stokes--Helfrich model and $\energy[\Sigma] = \energy[K] +\energy[T] + \energy[H]$ for the full model. The time axes are logarithmic to make the different evolutions comparable. In (a) the curves for $\energy[\Sigma]$ and $\energy[H]$ are indistinguishable.}
    \label{fig:energy_thick}
\end{figure}

We further quantify these changes in Fig. \ref{fig:energy_thick} by considering  the evolution of the energy ${\energy}_\Sigma$ and its various components, as well as the surface area $|\surf|$. Again ${\energy}_\Sigma$ decreases over time. In the current setting the evolution is dominated by $\energy[H]$. For the Surface Navier-Stokes-Helfrich model sudden changes in ${\energy}_\Sigma$ correspond to morphological changes, e.g. the transition to the three-fold-like shape at $t \approx 25$, and the transition towards the oblate shape at $t \approx 100$. For the full model also $\energy[T]$ contributes to the evolution, with an increase until $t \approx 2.5$, corresponding to the reduced thickness at the strongly negative curvature regions. The following decrease of $\energy[T]$ is associated with an increase of $\energy[H]$. This interplay between changes in thickness and shape evolution can also be seen in the evolution of the surface area $|\surf|$, which reaches a minimum at $t \approx 2.5$, before it increases again. The final shape is not only reached significantly faster in the full model, also the curves for ${\energy}_\Sigma$ qualitatively differ.

 \subsubsection{Coupling between tangential flow, shape changes and thickness variation}

\begin{figure}
    \centering
    \includegraphics[width=0.9\textwidth]{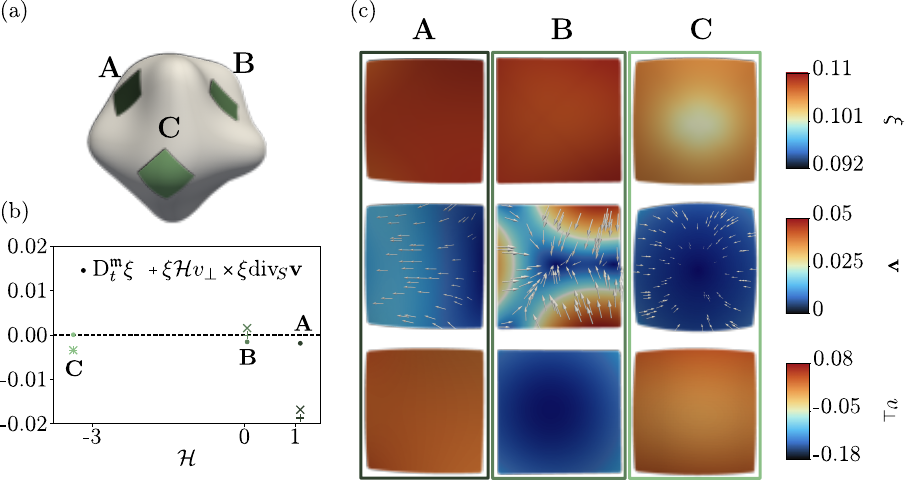}
    \caption{Local view on different quantities at a fixed time $t = 1.5$. (a) Full surface with chosen patches 'a', 'B' and 'C'. (b) Quantities of the average of the material time derivative of the thickness $\Dtm\height$ and the normal and tangential components $\height\meanCurv\veloN$ and $\height\divS\veloT$ of $\height\DivC\velo$ of these patches, shown with respect to the average mean curvature $\meanCurv$ of the patches. (c) Magnified patches. Shown in color are the thickness $\height$, the magnitude of the tangential velocity $\veloT$, and the normal velocity $\veloN$. For the tangential velocity we also show the direction by arrows, The length is scaled logarithmically with the magnitude.}
    \label{fig:panel_thick_2}
\end{figure}

 Having explored the qualitative changes compared with the reduced models of the shallow water model on a stationary surface and the Surface Navier-Stokes-Helfrich model, we finally aim to demonstrate that the proposed Surface Shallow-Water-Helfrich model maintains the solid-fluid duality and with it the tight coupling between tangent flow and shape changes in the presence of curvature and only enhances the evolution by an additional degree of freedom, which allows for area changes as a result of thickness variations. We consider the setting of the previous Section and analyze local features of the evolution. In Fig. \ref{fig:panel_thick_2} the solution at $t = 1.5$ is depicted. We select three representative patches, marked by 'A', 'B' and 'C' and color coded by various greens. We show $\xi$, $\veloT$ and $\veloN$ on these curved patches and compute the mean of $\meanCurv$ and the mean of all terms in the continuity equation \eqref{eq:conti}, i.e. $\Truess{\height} = \Dtm\height + \height \DivC\velo = \Dtm\height + \height \divS \veloT - \height \veloN \meanCurv = 0$. Fig. \ref{fig:panel_thick_2} shows the mean of $\Dtm\height$, $\height \divS \veloT$ and $\height \veloN \meanCurv$ in relation to the mean of $\meanCurv$ for the three patches. For 'B' the mean of $\meanCurv$ is essentially zero, shape changes (movement in normal direction) do not contribute to the continuity equation and changes in the height profile must be balanced by area growth/shrinkage. The corresponding values for $\Dtm\height$ and $\height \divS \veloT$ indicate this. In contrast for 'A' and 'C' the mean of $\meanCurv$ does not vanish, but changes in height are negligible. For these cases tangential flow, represented by $\height \divS \veloT$, and shape changes, represented by $\height \veloN \meanCurv$, are balanced, in accordance with the classical solid-fluid duality of the Surface Navier-Stokes-Helfrich model.

  \section{Conclusions} \label{sec:Conclusions}

  We have extended the Surface Navier-Stokes-Helfrich model by incorporating a thickness field in the spirit of a shallow-water equation. The thickness/height is measured with respect to the normal of the surface and variations in thickness lead to surface area growth/shrinkage. This replaces the inextensibility constraint of the Surface Navier-Stokes-Helfrich model by an incompressibility constraint of the thin film fluid. The model is derived by the Lagrange-d'Alembert principle. We develop a numerical algorithm to solve the equations, which extends known SFEM approaches for the Surface Navier-Stokes-Helfrich model, and demonstrate convergence and the physical implications of the model. The modification maintains the solid-fluid duality, inherent for the Surface Navier-Stokes-Helfrich model. The simulations indicate that the tight coupling between tangent flow and shape changes in the presence of curvature remains. However, the evolution is enhanced by an additional degree of freedom, which allows for area changes as a result of thickness variations. As a result, the dynamic is strongly enhanced and finds different pathways. 

  The model should be seen as a minimal model for fluid deformable surfaces,  accounting for variations in thickness in a consistent manner. While several parameters are accessible by measurements, e.g. the viscosity $\nu$, the friction parameter $\gamma$, or the bending rigidity $\kappa$, other parameters are less know. $a$ and $b$ are model parameters only chosen to favor a specific thickness, but have a strong impact on the evolution. This also holds for the specific value and the functional dependency of $\kappa(\xi)$. These parameters are chosen to highlight the coupling of the various terms. For quantitative simulations of epithelial tissue, e.g. in the motivating examples in morphogenesis, these coupling terms need to be calibrated.

\section*{Author Contributions}

A.V. designed the research. J.M.S. derived the model, developed and implemented the algorithms and carried out all simulations and analyzed the data. J.M.S. and A.V. wrote the article.

\section*{Acknowledgments}

This work was supported by the German Research Foundation (DFG) through the research unit FOR3013, ``Vector- and Tensor-Valued Surface PDEs,'' within project TP01, ``Numerical methods for surface fluids'' (project number VO 899/29-2). We acknowledge computing resources provided by JSC within project MORPH and by ZIH within project WIR.

\section*{Data Availability}

Data will be made available on request.

  \section*{Appendix} 
  \appendix
  \section{Variational derivatives} \label{app:Derivations}
  In this section we show the derivations of the model described in Section \ref{sec:Model Derivation} in detail. For a more detailed description of the differential operators and the corresponding differentiation and integration rules we refer to \cite{bachini2023derivation, nitschke2025beris}. Several terms have already been considered in \cite{bachini2023derivation} and are here listed only for convenience. We start with deriving the variational derivatives of the potential energy $\energy[P] = \energy[T] + \energy[H]$. The deformation derivative $\deform$ in direction of $\testVelo\in\TRS$, is defined as
  \begin{align}
      \deform f&\coloneqq \left.\frac{\operatorname{d}}{\operatorname{d}\varepsilon}\right\vert_{\varepsilon = 0} f\vert_{\para + \varepsilon\testVelo}&\forall f\in\TS, & \quad\testVelo\in\TRS.
  \end{align}
  By \cite{nitschke2023tangential}, it holds $\deform\metric_{ij} = -\left((\idS\,\gradC\testVelo)_{ij} + (\idS\,\gradC\testVelo)_{ij}^T\right)$, where $\metric_{ij}[\para] = (\partial_i\para, \partial_j\para)$ are the components of the first fundamental form. With the symmetry of the metric tensor, choosing $\deform \height = -\height\DivC\testVelo$ and by $\deform\partial_i = \partial_i\deform$, we get 
  \begin{align}\label{eq:deform_scalar}
      \deform \|\gradS \height \|^2 =& \deform\left(\metric^{ij}\partial_i \height\partial_j \height\right) \nonumber\\
      =& \deform\metric^{ij}\partial_i \height \partial_j \height 
        + 2\metric^{ij}\deform\partial_i \height \partial_j\height \nonumber\\
      =& - 2\gradS\height\otimes\gradS\height:\gradC\testVelo - 2\metric^{ij}\partial_i\height\partial_j(\height\DivC\testVelo) \nonumber\\
      =& - 2\gradS\height\otimes\gradS\height:\gradC\testVelo - 2 \gradS\height\cdot\gradS\left(\height\DivC\testVelo\right)\,.
  \end{align} 
  Thus it holds
  \begin{align}
      \LGrad{\frac{\delta \energy[T]}{\delta\para}}{\testVelo} =& \,\int_\surf\frac{\heightparamI}{2} \deform\|\gradS \height\|^ 2  \nonumber\\ 
      & \,\phantom{\int_\surf}+ \frac{\heightparamII}{2}\deform(\height-\height_0)^2+\left(\frac{\heightparamI}{2} \|\gradS \height\|^ 2 + \frac{\heightparamII}{2}(\height-\height_0)^2\right)\DivC\testVelo\,d\surf
      \nonumber\\
      =& \,\int_\surf - \heightparamI\gradS\height\otimes\gradS\height : \gradC\testVelo -  \heightparamI\gradS\height \cdot \gradS\left(\height\DivC\testVelo\right) 
      \nonumber\\
       & \,\phantom{\int_\surf}  - \heightparamII (\height-\height_0)\height\DivC\testVelo
       + \left(\frac{\heightparamI}{2} \|\gradS \height\|^ 2 + \frac{\heightparamII}{2}(\height-\height_0)^2\right)\DivC\testVelo \,d\surf
      \nonumber\\
      =& \, \LGrad{- \heightparamI \gradS\height\otimes\gradS\height + \heightparamI \height\lapS\height\idS}{\gradC\testVelo}
      - \LGrad{ \heightparamII (\height-\height_0)\height\idS}{\gradC \testVelo} \nonumber\\
      &+ \LGrad{\left(\frac{\heightparamI}{2} \|\gradS \height\|^ 2 + \frac{\heightparamII}{2}(\height-\height_0)^2\right)\idS}{\gradC\testVelo}
      \nonumber\\
      =&\LGrad{\DivC\left(\heightparamI\gradS\height\otimes\gradS\height - \left(\heightparamI\height\lapS\height-\heightparamII(\height-\height_0)\height\right)\idS\right)}{\testVelo}\nonumber\\
      &-\LGrad{\DivC\left.\left(\frac{\heightparamI}{2} \|\gradS \height\|^ 2 + \frac{\heightparamII}{2}(\height-\height_0)^2\right)\idS\right)}{\testVelo}\,.
  \end{align}
  According to \cite{bachini2023derivation} and with $\deform\height = -\height\DivC \testVelo$, $\deform\meanCurv = \divS(\normal\gradC\testVelo) - \shapeOp\colon\!\gradC\testVelo$ it holds
  \begin{align}
      \LGrad{\frac{\delta\energy[H]}{\delta\para}}{\testVelo} 
      =& \int_\surf \deform\left(\frac{\bendStiff(\height)}{2}\meanCurv^2\right) 
      + \frac{\bendStiff(\height)}{2}\meanCurv^2\DivC\testVelo \,d\surf\nonumber\\
      =& \int_\surf \frac{\bendStiff'(\height)}{2}\deform\height\meanCurv^2 + \bendStiff(\height)\meanCurv\deform\meanCurv+ \frac{\bendStiff(\height)}{2}\meanCurv^2\DivC\testVelo\,d\surf\nonumber\\
      =&\LGrad{\frac{\bendStiff(\height)-\bendStiff'(\height)\height}{2}\meanCurv^2}{\DivC \testVelo}-\LGrad{\normal \otimes \gradS(\kappa(\height)\meanCurv)+\kappa(\height)\meanCurv\shapeOp}{\gradC\testVelo}\nonumber\\
      =&
      \LGrad{\divC\left(\bendStiff(\height)\meanCurv\shapeOp-\frac{\bendStiff(\height)-\bendStiff'(\height)\height}{2}\meanCurv^2\idS + \normal\otimes\gradS(\bendStiff(\height)\meanCurv)\right)}{\testVelo}\,.
  \end{align}
  Using $\divS\shapeOp = \gradS\meanCurv$, $\gradS\idS = 0$ and $\gradS\bendStiff(\height) = \bendStiff'(\height) \gradS\height$, the normal and tangential parts of the bending forces read
  \begin{align}
  \normal\D[\para]\energy[H] &=
      \shapeOp : \left(\bendStiff(\height)\meanCurv\shapeOp - \frac{\bendStiff(\height)-\bendStiff'(\height)\height}{2}\meanCurv^2\idS\right)+\divS\gradS(\bendStiff(\height)\meanCurv)\nonumber\\
      &= \bendStiff(\height)\meanCurv\|\shapeOp\|^2 - \frac{\bendStiff(\height)-\bendStiff'(\height)\height}{2}\meanCurv^3
         + \lapS(\bendStiff(\height)\meanCurv)\\
      \idS\D[\para]\energy[H] &= \divS\left(\bendStiff(\height)\meanCurv\shapeOp -\frac{\bendStiff(\height)-\bendStiff'(\height)\height}{2}\meanCurv^2\idS\right) - 
      \shapeOp\gradS(\bendStiff(\height)\meanCurv)\nonumber\\
      &= \gradS(\bendStiff(\height)\meanCurv)\shapeOp 
         + \bendStiff(\height)\meanCurv \gradS\meanCurv - \gradS\left(\frac{\bendStiff(\height)-\bendStiff'(\height)\height}{2}\right)\meanCurv^2 \nonumber\\
      &\phantom{=}- (\bendStiff(\height)-\bendStiff'(\height)\height)\meanCurv\gradS\meanCurv - 
      \gradS(\bendStiff(\height)\meanCurv)\shapeOp \nonumber\\
      &=\frac{\bendStiff''(\height)\height\gradS\height}{2}\meanCurv^2 + \bendStiff'(\height)\height\meanCurv\gradS\meanCurv\,.
  \end{align}
  Variations with respect to the thickness $\height$ lead to
  \begin{align}\label{eq:dFTh}
      \LGrad{\frac{\delta \energy[T]}{\delta \height}}{\testHeight} =& \LGrad{-\heightparamI\lapS\height + \heightparamII(\height-\height_0)}{\testHeight}
  \end{align}
  and 
  \begin{align}\label{eq:dFHh}
      \LGrad{\frac{\delta \energy[H]}{\delta \height}}{\testHeight} =& \LGrad{ \frac{\bendStiff'(\height)}{2}\meanCurv^2}{\testHeight}\,.
  \end{align}
  Next, we consider the dissipative forces. Similar to \cite{bachini2023derivation} it holds
  \begin{align}
      \LGrad{\frac{\diss[V]}{\delta \velo}}{\testVelo} = \LGrad{\visc\divC(\height\deformTensor)}{\testVelo}\,,
  \end{align}
  According to \cite{bachini2023derivation} it holds
  \begin{align}
      \LGrad{\frac{\delta\diss[R]}{\delta\velo}}{\testVelo} =& \LGrad{\friction\velo}{\testVelo}
  \end{align}
  and with $\deform\height = -\height\DivC\testVelo$, it holds
  \begin{align}
      \LGrad{\frac{\delta{\diss}_\height}{\delta\velo}}{\testVelo} &= \int_\surf \mobiH\Truess{\height}\deform\Truess{\height}  \,d\surf = 0\,. 
  \end{align}
  For an arbitrary functional $\mathfrak{F} = \sint{f}$ with $f\in\TS$, by \cite{nitschke2025beris} it holds 
  \begin{equation*}
      \LGrad{\frac{\delta\mathfrak{F}}{\delta\velo}}{\testVelo} = \sint{\veloDer f},
  \end{equation*}
  where $\veloDer$ is the velocity derivative in direction of $\testVelo\in\TRS$, defined by
  \begin{equation}
      \veloDer f \coloneqq \left.\frac{\operatorname{d}}{\operatorname{d}\varepsilon}\right\vert_{\varepsilon=0}f[\velo + \varepsilon\testVelo].
  \end{equation}
  By $\veloDer \Truess{\height} = \deform \height + \height\DivC\testVelo$ (c.f. \cite{nitschke2025beris}) and $\deform\height = -\height\DivC\testVelo$, it holds
  \begin{equation} \label{eq:dRpV}
      \LGrad{\frac{\delta\con[p]}{\delta\velo}}{\testVelo} = 
            \sint{\press(\deform\height +\height\DivC\testVelo)}=0\,.
  \end{equation}
  For the volume constraint we get 
  \begin{equation}\label{eq:RVolV}
     \LGrad{\frac{\delta\diss[Vol]}{\delta\velo}}{\testVelo} = 
     \LGrad{\lagrVol\normal}{\testVelo} \,.
  \end{equation}
  Considering the rate of change of the thickness $\Truess\height$, simple calculations show that
  \begin{align}\label{eq:dRVh}
      \LGrad{\frac{\delta\diss[V]}{\delta \Truess{\height}}}{\testHeight} &= 0\,,\\ \label{eq:dRRh} 
      \LGrad{\frac{\delta\diss[R]}{\delta\Truess{\height}}}{\testHeight} &= 0\,,\\ \label{eq:dRhh}
      \LGrad{\frac{\delta{\diss}_{\height}}{\delta\Truess{\height}}}{\testHeight} &= \LGrad{\mobiH \Truess{\height}}{\testHeight}\,, \\ \label{eq:dRph}
      \LGrad{\frac{\delta\con[p]}{\delta\Truess\height}}{\testHeight} &= 
            \LGrad{\press}{\testHeight}\,,\\\label{eq:dRVolh}
      \LGrad{\frac{\delta\diss[Vol]}{\delta\Truess\height}}{\testHeight} &= 0\,.
  \end{align}
  
  The only non-trivial variational derivatives with respect to the Lagrange function and Lagrange multiplier read
  \begin{align}\label{eq:dRpp}
      \LGrad{\frac{\delta\con[p]}{\delta\press}}{\testPress} &= 
            \LGrad{\Truess{\height}}{\testPress}\,, \\\label{eq:dRVoll}
      \frac{\partial \diss[Vol]}{\partial \lagrVol} &=\sint{\velo \cdot \normal} \,.
  \end{align}
  Note that $\press$ only appears explicitly in \eqref{eq:dRph}. Substituting \eqref{eq:dFTh}, \eqref{eq:dFHh} and \eqref{eq:dRVh}-\eqref{eq:dRVolh} into \eqref{eq:Lagrange-DAlembert}, we get 
  \begin{align}\label{eq:thickness_dynamic}
      \mobiH \Truess{\height} = \heightparamI\lapS\height - \heightparamII(\height-\height_0) -\frac{\bendStiff'(\height)}{2}\meanCurv^2-\press \,.
  \end{align}
  By \eqref{eq:dRpp} it holds $\mobiH \Truess{\height} = 0$. Therefore, equation \eqref{eq:thickness_dynamic} only serves as an explicit description of $\press$, and we can eliminate the equation alongside the Langrange function from the system.
  
  In Section \ref{sec:Surface_NSH}, there is an additional constraint of the thickness ${\diss}_\height$. By \cite{nitschke2025beris} and $\deform\height = -\height\DivC\testVelo$, the variation with respect to $\velo$ reads 
  \begin{equation}
    \LGrad{\frac{\delta {\diss}_\height}{\delta \velo}}{\testVelo} = \sint{\lagrHeight\deform \height} = \sint{ -\lagrHeight\height\DivC \testVelo} = \sint{\left(\gradS\left(\lagrHeight\height\right) + \lagrHeight\height\meanCurv\normal\right)\testVelo} \,.
  \end{equation} 
  
  \bibliographystyle{unsrt}
  \bibliography{bib}
\end{document}